\documentstyle[aps,prb,twocolumn]{revtex}

\begin{document}
\draft
\title{Phase Diagram of Multilayer Magnetic Structures}
\author{V.D. Levchenko}
\address{M.V. Keldysh Institute for Applied Mathematics, Russian Academy of Sciences}
\author{A.I. Morosov and A.S. Sigov}
\address{Moscow State Institute of Radioengineering, Electronics, and Automation (Technical University)}
\date{\today}
\maketitle
\begin{abstract}
\end{abstract}

\maketitle
\begin{abstract}
Multilayer ``ferromagnet{}--{}layered antiferromagnet'' (Fe/Cr) structures frustrated
due to the roughness of layer interfaces are studied by numerical modeling methods.
The ``thickness{}--{}roughness'' phase diagrams for the case of thin ferromagnetic film on
the surface of bulk antiferromagnet and for two ferromagnetic layers separated by an
antiferromagnetic interlayer are obtained and the order parameter distributions for all
phases are found. The phase transitions nature in such systems is considered. The
range of applicability for the ``magnetic proximity model'' proposed by Slonczewski is
evaluated.
\end{abstract}
\section{Introduction}
Multilayer magnetic structures have attracted a widespread attention after
discovery of the phenomenon of giant magnetoresistance.\cite{ref1} Within the span of
eleven years a great many of papers devoted to the multilayer structures has been
published. Considerable recent attention has been focused on multilayer
``ferromagnet{}--{}layered antiferromagnet'' structures. An example of such structures,
according to recent experiments on neutron diffraction, is given by multilayer Fe/Cr
structures,\cite{ref2,ref3} in which antiferromagnetic chromium layers of the thickness $d<4.5 nm$
consist of ferromagnetic atomic planes with antiparallel orientation of spins in adjacent
planes. The spins of chromium atoms lie in these planes, the plains in their turn being
parallel, on the average, to the layer interfaces. An analogous structure has been
observed in manganese layers in multilayer Fe/Mn structures.\cite{ref4,ref5}

In multilayer ``ferromagnet{}--{}layered antiferromagnet'' structures, the exchange
interaction between ferromagnetic layers is caused mainly by the interaction via
antiferromagnetic order parameter, whereas the RKKY interaction far away from the
Neel temperature provides only a small additive to the first one.\cite{ref6} To describe the
interaction via antiferromagnetic order parameter, Slonczewski proposed a
phenomenological model of ``magnetic proximity'' in whose context the layers of
ferromagnet are assumed to be magnetized practically uniformly and considerable
distortions of the order parameter are supposed to take place only in the layers of
antiferromagnet.\cite{ref7}

The presence of atomic steps at layer interfaces that change the local thickness
of the antiferromagnet by one monoatomic layer gives rise to frustrations in the system
(Fig.~\ref{Fig1}a). In this case the uniform distribution of the order parameters in the layers does
no longer correspond to the energy minimum.

If the characteristic distance between atomic steps at the layer interface (we
shall call it the step width $R$) exceeds some critical value, it becomes energetically
favorable to break the layers up into domains \cite{ref8,ref9} (Fig.~\ref{Fig1}b). 
The domain boundaries in
layer planes coincide with the atomic step edges.
It should be noted that the magnitude of $R$ strongly depends on the technology conditions
\cite{Schmidt:1999}.

Recent investigations on the state of ferromagnetic iron film on the rough (001)
chromium surface have shown the presence of several magnetic phases depending
upon the film thickness and roughness rate (the value of $R$).\cite{ref10}

All the aforesaid bears witness to considering the problem of
``thickness{}-{}roughness'' phase diagram of ``ferromagnet{}--{}layered antiferromagnet'' magnetic
structures as a high priority task.

The paper has the following structure. The consideration begins with a simple
model that allows one to describe the behavior of the system in question qualitatively.
In the third section one can find model results on the phase diagram of the two-layer
system: thin ferromagnetic film on antiferromagnetic substrate (or vice versa). The
phase diagram for ``ferromagnet{}--{}antiferromagnet{}--{}ferromagnet'' three-layer structure
is found and discussed in the fourth section, the treatment being possible to be
generalized to multilayer. In conclusion the main results are adduced.
\section{Description of the model}
When describing the multilayer structure, we restrict ourselves to the mean-field
approximation. Let us introduce the order parameter for each magnetic layer: the
magnetization vector for the layers of ferromagnet and antiferromagnetism vector,
equal to the difference in magnetization vectors of the sublattices, for the layers of
antiferromagnet.

It is known that in sufficiently thin magnetic layers with the thickness of some
nanometers, the atomic spins lie in the layer planes. Therefore, at $T<T_C$, $T_N$ (here $T_C$ is
the Curie temperature for single ferromagnetic layers, and $T_N$ is the Neel temperature
of antiferromagnetic layers), it is possible to describe the local value of the order
parameter in the layer planes by the angle $\theta$ of the order parameter vector with a
certain given axis in the layer plane. The order parameter modulus is assumed to
remain unchanged in each layer.

In the framework of the assumptions made above, the exchange energy $W_i$ caused by a
nonuniformity within the $i-$th layer can be introduced as
\begin{equation}
W_i = \frac{J_i S_i^2}{2 b_i} \int\left(\nabla\theta_i\right)^2\, d V,
\label{eqn1}
\end{equation}
where the integral is taken over the $i-$th layer volume, $J_i$ is the exchange stiffness, $S_i$ is
the mean spin of the atom, $b_i$ is the inter-atomic distance.

By varying Eq.~(\ref{eqn1}) with respect to $\theta_i$ we get the equation for the order parameter
distribution in the layer:
\begin{equation}
 \Delta \theta_i = 0.
\label{eqn2}
\end{equation}
To obtain the true boundary conditions one needs a more exhaustive procedure.
Owing to frustrations arising at the layers boundaries the difference
 $\theta_i-\theta_{i+1}$ can become 
large enough at the interlayer boundary, whereas inside the layers the
 frustrations are 
absent, the value of $\theta_i$ varies smoothly, and the difference of $\theta_i$ 
values in the nearest 
cells is small. That is why, when calculating exchange energy within the X-Y
model, we may expand cosine of the difference of the angles $\theta_i$ 
in neighboring cells in 
power series if the cells belong to the same layer, but we'll not do that if
 the cells belong 
to different layers. By differentiating the energy with respect to the quantity
$\theta_i$ in the 
cell lying in the layer boundary, 
we come to the equation which takes the following form 
as one passes to the continuous representation:
\begin{equation}
\Delta \theta_i - \frac{\partial\theta_i}{\partial n} =
 \pm \frac{J_{f,af} S_{i+1}}{J_i S_i} \sin\left(\theta_i - \theta_{i+1}\right),
\label{eqn3}
\end{equation}
where $\Delta$ is the two-dimensional Laplacian in the layer plane,
$\frac{\partial}{\partial n}$ is the derivative in the
direction of the outer normal to the interface plane, the exchange constant $J_{f,af}$
describes the interaction of spins corresponding to different layers, and all distances
are normalized to the value of $b_i =b$ which is assumed to be the same in all layers. The
signs in the right-hand part of Eq.~(\ref{eqn3}) are opposite for different sides of the atomic step
at the interface. For the free surface of the layer the right-hand side of Eq.~(\ref{eqn3}) equals
zero.

The energy of exchange interaction between adjacent layers takes the from
\begin{equation}
 W_{i,i+1} =
 \pm \frac{J_{f,af} S_i  S_{i+1}}{b^2} \int\cos\left(\theta_i - \theta_{i+1}\right)\,d S,
\label{eqn4}
\end{equation}
where the numbers $i$ and $i+1$ denote corresponding layers, the integration is performed
over the interface of the layers, and the sign in the right member coincides with that
one in Eq.~(\ref{eqn3}).

By varying the energy of interlayer interaction represented in terms of the
continuum model distribution one gets an equation similar to Eq.~(\ref{eqn3}) but without the
first summand in the left-hand side. This renders it impossible to pass from Eq.~(\ref{eqn3}) to
Eq.~(\ref{eqn2}) if adjacent layers are identical in composition.

 The atomic steps divide the whole interlayer surface into regions of two types: in the
regions of the first type the surface energy takes its minimum for $\theta_i=\theta_{i+1}$ and in the
regions of the second type the energy is a minimum for $\theta_i = \pi - \theta_{i+1}$,

To find the order parameters distributions in the multilayer structure it is
necessary to solve the system of differential equations (\ref{eqn2}) with boundary conditions
(\ref{eqn3}).

Now let us take up the applicability of this simple model to real multilayer
structures. The continuum approximation is valid only in the case that characteristic
measures of the problem are much higher than the interatomic distance. The layer
thicknesses in multilayer structures as well as the distances between atomic steps are
of about several nanometers, so they may be considered as far exceeding the lattice
parameter. Hence the continuum approximation is applicable for estimations and
qualitative treatment.

In the model proposed, the exchange interaction is assumed to be isotropic, i.e.
identical in the layer plane and in perpendicular direction. The model with anisotropic
exchange interaction is reduced to the given one by renormalization of the distance
scale in any of two non-equivalent directions.

An interdiffusion of atoms from neighboring layers leads only to renormalization
of $J_{f,af}$, if the region of mutual stirring extends up to several monoatomic layers, i.e. is
of atomic size scale. This constant value is obtained by microscopic calculations.\cite{ref11}

Eqs. (\ref{eqn2}), (\ref{eqn3}) are written in the exchange approximation, but can be easily
generalized to the case of small anisotropy in the layer plane.

\section{The method of calculations}
Let the atomic step edges be rectilinear and parallel to each other. The $x$-axis of
the coordinate system lies in the layer plane and is perpendicular to the step edges, and
the $z$-axis is perpendicular to the layer plane (two-dimensional case). 

The assumed set of equations includes Laplace equations (\ref{eqn2}) for the each plane 
layer $-\infty<x<\infty,\ 0<z<a_i$, where $i=1\dots n$ is the layer number
and $a_i$ is the layer width, the boundary conditions (\ref{eqn3}) at interlayer 
surfaces being nonlinear and discontinuous. For numerical calculations we
reduce this set to the system of unidimensional integral equations.

We require the function $\theta_i(x,z)$ to be continuous inside the region
$0<x<L,\ 0<z<a_i$ (here $L$ is half of the modeling region length along $x$ axis), 
and additionally
$\left.\frac{\partial \theta_i}{\partial x}\right|_{x\to 0,L}\to 0$.
To make the function  $\theta_i(x,z)$ periodic with the period $2L$ we also
put $\theta_i(-x,z) = \theta_i(x,z)$, $-L < x \leq L$. Let us put a uniform 
net $\{x_j\}$ with cells number $N$ and step $\Delta^x=2 L/N$
on the interval $-L < x \leq L$ and
introduce the function $\theta_i(x,z)$, as well as the right-hand part of 
Eq.~(\ref{eqn3}) as the Fourier series:
\begin{eqnarray}
\theta_i(x,z) =
 \Phi_{0,i}(z) + \sum\limits_{k=1}^N \Phi_{k,i}(z) e^{i\frac{\pi}{L} kx},
\label{SolFourie}\\
\sigma^\pm_i(x)\sin(\delta\theta^\pm_i(x)) =
 \Psi^{\pm}_{0,i} + \sum\limits_{k=1}^N \Psi^{\pm}_{k,i} e^{i\frac{\pi}{L} kx}.
\label{BCFourie}
\end{eqnarray}
Here $\sigma^\pm_i(x) = \pm \frac{J_{f,af} S_{i\pm   1}}{J_i S_i}$
is a stepwise function changing its value at the step boundaries and 
$$
\delta\theta_i \equiv \delta\theta^+_i(x) = -\delta\theta^-_{i+1}(x)
 = \theta_{i+1}|_{z=0}-\theta_i|_{z=a_i}.
$$
Substituting Eqs.~(\ref{SolFourie}-\ref{BCFourie}) into the starting
equations one comes to normal differential equations for any harmonic 
$k=1,2\dots N$ inside each layer $i$:
\begin{equation}
\frac{d^2 \Phi_k}{d z^2} - A_k^2 \Phi_k = 0, \label{LaplaceFTx}
\end{equation}
(the index $i$ is omitted here and further, where it does not make 
understanding difficult) with the following boundary conditions
\begin{equation}
\left(\frac{d \Phi_k}{d z} \pm A_k^2 \Phi_k\right)_{z=a,0} = -\Psi^{\pm}_k,
 \label{BCFTx}
\end{equation}
$$
A_k = \frac1{\Delta^x}\sqrt{1-\cos\left(\frac{\pi}{L}k\Delta^x\right)}.
$$
The solution of Eq.~(\ref{LaplaceFTx}) is sought in the form
$$\Phi_k(z)=C_1^k e^{A_k z} + C_2^k e^{-A_k z}$$ 
with the constants $C_1^k$, $C_2^k$ 
derived from boundary conditions (\ref{BCFTx}). 
As a result one has
\begin{equation}
\Phi_k(z) = K_k^+(z) \Psi^+_k + K_k^-(z) \Psi^-_k,\label{ZSolution}
\end{equation}
where
\begin{eqnarray}
K_k^+(z) = dK_k e^{-(a-z)A_k}\left[(1+A_k)+(1-A_k)e^{-2zA_k}\right]\nonumber\\
K_k^-(z) = dK_k e^{-zA_k}\left[(1+A_k)+(1-A_k)e^{-2(a-z)A_k}\right]\nonumber\\
dK_k = - A_k^{{-}1} \left[(1+A_k)^2 - (1-A_k)^2 e^{-2 a A_k}\right]^{-1}.\nonumber
\end{eqnarray}
The equation for zeroth--order Fourier coefficients has the form
$$\frac{d^2 \Phi_0}{d z^2} = 0$$
with boundary conditions
$$\left.\frac{d \Phi_0}{d z}\right|_{z=a,0} = -\Psi_0^\pm.$$
With account of $\frac{d \Phi_0}{d z} = 0$ for free boundaries of the top
and bottom layers, and also of the constant ratio
$$
\Psi_{0,i-1}^+(\delta\theta^+_{i-1}) / \Psi_{0,i}^-(\delta\theta^-_{i})
 = const
$$
for the neighboring layers, one has $\Psi_0^\pm = 0$.
If one separates out the average angle 
$$\overline {\delta\theta} = \frac1{2L}\int_{-L}^L \delta\theta(x) dx$$
and the variation
$$\widetilde {\delta\theta}(x) = \delta\theta(x) - \overline {\delta\theta},$$
it is possible to find
\begin{eqnarray}\nonumber
0 &=& \int_{-L}^L \sigma(x) 
  \sin (\widetilde{\delta\theta}(x) + \overline{\delta\theta}) dx\\\nonumber
  &=& \cos\overline{\delta\theta}
\int_{-L}^L\sigma(x)\sin\widetilde{\delta\theta}(x) dx  \\\nonumber
  &+& \sin\overline{\delta\theta}
\int_{-L}^L\sigma(x)\cos\widetilde{\delta\theta}(x) dx.
\end{eqnarray}
Hence the average angle is
\begin{equation}
\overline {\delta\theta} = n\pi - \arctan \left(
  \frac{\int_{-L}^L\sigma(x)\sin\widetilde{\delta\theta}(x) dx}
       {\int_{-L}^L\sigma(x)\cos\widetilde{\delta\theta}(x) dx}
 \right). \label{AveAngle}
\end{equation}
With Eqs~(\ref{SolFourie}-\ref{BCFourie}) and 
(\ref{ZSolution}-\ref{AveAngle}) one can obtain the desired integral 
equation for the function $\delta\theta_i$:
\begin{eqnarray}\nonumber
\delta\theta_i(x) =\overline{\delta\theta_i}
 + \sum\limits_{k=1}^N e^{i\frac{\pi}{L} kx} \frac1{2L}
 \int_{-L}^L e^{-i\frac{\pi}{L} k\xi} d\xi \\\nonumber \times\Bigl[
 K^-_{i{+}1,k}(0)\sigma^-_{i{+}1}\sin\delta\theta_i
+K^+_{i{+}1,k}(0)\sigma^+_{i{+}1}\sin\delta\theta_{i+1} \\\nonumber
-K^-_{i,k}(a_i)\sigma^-_i\sin\delta\theta_{i{-}1}
-K^+_{i,k}(a_i)\sigma^+_i\sin\delta\theta_i\Bigr]
\\
 \equiv \hat I (K,\delta\theta_i,\delta\theta_{i\pm1}).\nonumber
\end{eqnarray}
To solve this equation numerically we use a simple iteration procedure
$$
\delta\theta_i^{n+1} = (1{-}F_i(x))\ \delta\theta_i^{n}
 + F_i(x)\ \hat I (F_{k,i}K,\ \delta\theta_i^n,\ \delta\theta_{i\pm1}^n)
$$
with $0 < F(x),F_k \leq 1$ which are filters providing stability of the 
procedure and increasing its convergence rate. 
The iteration procedure is over when the residual solution
$$
\varepsilon = \max\left|
 \delta\theta_i^{n}(x) - \hat I (\delta\theta_i^n,\delta\theta_{i\pm1}^n)
\right|
$$
becomes less than a given value (normally, $\varepsilon < 10^{-6}$).

Then the solution over the whole region can be reconstructed from the formula:
\begin{eqnarray}\nonumber
\theta_i(x,z) =\overline{\delta\theta_i}
 + \sum\limits_{k=1}^N e^{i\frac{\pi}{L} kx} \frac1{2L}
 \int_{-L}^L e^{-i\frac{\pi}{L} k\xi} d\xi \\\nonumber \times\Bigl[
 K^-_{i,k}(z)\sigma^-_i\sin\delta\theta_{i{-}1}
+K^+_{i,k}(z)\sigma^+_i\sin\delta\theta_i\Bigr]
\end{eqnarray}

As an initial approximation, one can take the function of the form
$$\delta\theta_i^{0}(x) = \pi\sum_j\pm\eta(x-x_j),$$
where $\eta(x-x_j)$ is a unit stepwise function with a jump at the
position of crystal lattice defect location $x_j$ and the sign $\pm$ means 
that the given function can enter the sum with both plus and minus signs.
As a result, one has $\sim 2^{N_j}$ possible initial approximations, where
$N_j$ is the number of defects in the region under calculations. It should
be noted that to each of the initial conditions there can correspond one 
of the solutions of initial nonlinear equation 
(a local minimum of potential energy). To find the global minimum, it should 
be compared the energies corresponding to all of the solutions obtained.

The solution depends on the values of $a_i$, characteristic
distance $R$ between the edges of the steps, and the parameter
\begin{equation}
\alpha_f = \frac{J_{f,af} S_{af}}{J_f S_f}
\label{eqn5}
\end{equation}
characterizing the ratio of the exchange interaction energies of the nearest spins
belonging to different layers and to ferromagnetic layer, respectively,
and also the parameter $\alpha_{af}$ defined by Eq.(\ref{eqn5}) with the subscript
substitution: $f \leftrightarrow af$.

\section{Thin ferromagnet film on antiferromagnet}
Let us consider a thin ferromagnet film of the thickness $a$ (in normalized
unitless scale) on the surface of layered antiferromagnet. The problem of
antiferromagnet film on ferromagnetic substrate is easily reduced to the given one.

Three different states of the film-substrate system are possible, depending on the
relationship between indicated parameters. If the atomic steps on the film-substrate
interface are well away one from another, the frustrations give rise to domain
formation.\cite{ref8,ref9} The magnetization orientation in each domain is conditioned by the
surface energy minimum (Fig.~\ref{Fig1}b). Hence the magnetizations in the neighboring
domains are antiparallel. The structure of Neel walls demarcating the domains depends
on the value of dimensionless parameter $\gamma$ equal to the ratio of exchange energies in
the film and in the substrate:
\begin{equation}
\gamma = \frac{J_f S_f^2}{J_{af} S_{af}^2} = \frac{\alpha_{af}}{\alpha_{f}}.
\label{eqn6}
\end{equation}
We have already considered the case where the order parameter distortions in
the substrate are negligible.\cite{ref12} Such an approximation conforms to $\gamma \ll 1$. A
distinguishing feature of the domain wall arising is an increase in its width with a
distance from the edge of the atomic step giving rise to the wall.
The parameter $\alpha_f a$ therewith plays an important part in the wall width behavior.
The width of the domain wall $\delta(z)$ is interpreted as the distance between the points with
coordinates $(x_1,z)$ and $(x_2,z)$, which correspond to $\theta_1= \pi /4$ 
and $\theta_2=3 \pi /4$, respectively.

If $\alpha_f a \ll 1$, the wall width $\delta$ dependence on $z$ is of no significance and one has
a one-dimensional problem. The wall width magnitude $\delta_f$ can be estimated from 
the following simple consideration.

The $|\nabla \theta|$ value inside the wall is of the order of $\delta_f^{-1}$. 
Estimating the energy $W_1$ (Eq.~\ref{eqn1}) per the domain wall unit length, one obtains
\begin{equation}\label{w1}
w_1 \approx \frac{J_f S_f^2 a}{b \delta_f}.
\end{equation}
At the same time, the spins on the interface are 
frustrated in the region $|x|\lesssim \delta_f$,
resulting in the energy increase in excess of the minimum by the value
\begin{equation}\label{w2}
w_2 \approx \frac{J_{f,af} S_f S_{af}\delta_f}{b}
\end{equation}
per the domain wall unit length.

Minimizing the sum $w=w_1+w_2$ one finds
\begin{equation}
 \delta_f \approx \sqrt{a / \alpha_f},
\label{eqn7}
\end{equation}
and the domain wall energy per unit length is
\begin{equation}
w \approx \frac{J_f S_f^2}{b} \sqrt{a \alpha_f} \sim \frac{S_f}{b} \sqrt{a J_f J_{f,af} S_f S_{af}}.
\label{eqn8}
\end{equation}
The precise numerical calculations of $\delta_f$ and $w$ values for a wide range
of parameters $\alpha_f$ and $a$ confirm the estimations given above 
(the same relates to all cases considered below).

If $\alpha_f a \gg 1$, the domain wall width essentially increases as compared to the
value of $\delta_0^f$ at the interface. 
The $\delta_f(z)$ characteristic
dependence is shown in Fig.~\ref{Fig2}.
The increase is practically linear (${\partial\delta}/{\partial z}\approx 1$), with
slowing down nearly to zero at a free surface of the film.

One can find the values of $\delta_0^f$ and $w$ using the approach analogous 
to that proposed for the case $\alpha_{f} a \ll 1$. In the region $\delta_0^f \ll \rho \ll a$
the value of $|\nabla \theta|$ is proportional to $\rho^{-1}$, where $\rho$ is
the distance from the step in the $xz$ plane, just similar to the 
Kosterlitz-Thouless vortex \cite{Kosterlitz:1973}. 
The value $w_1$ is equal to
\begin{eqnarray}
 w_1 \approx \frac{\pi}{2} \frac{J_f S_f^2}{b} \ln\left( a/\delta_0^f \right),
\end{eqnarray}
and $w_2$ is given by Eq.~\ref{w2} with substitution of $\delta_0^f$ instead of $\delta_f$.
After minimization one has
\begin{eqnarray}
 \delta_0^f \approx \frac{1+\alpha_f}{\alpha_f},
\label{eqn9}\\
 \delta_f(a/2) \approx a,
\label{eqn10}
\end{eqnarray}
and $w \approx w_1 \gg w_2$.

 The value of $\delta_0^f$ is of the order of interatomic distance, and a mean wall width
comprises tens of angstroms, i.e. the domain walls resulting from frustrations are much
more narrow than ``normal'' domain walls in ferromagnets whose width is due to a
competition between the exchange and anisotropy energies.

In the case of iron film on chromium substrate considered in this paper, the
value of $\gamma$ is high, $\gamma \gg 1$. For such conditions the domain wall structure is more
complicated because the order parameter distortions can extend to the substrate as well.

If $\alpha_{af}\gamma a \ll 1$, the order parameter distortions in the
substrate are small and the domain wall characteristics do not differ from those in the
case $\gamma \ll 1$, $\alpha_f a \ll 1$.

For $\alpha_{af}\gamma a \gg 1$, as one can see from the results of modeling
(Fig.~\ref{Fig3}),
 two characteristic lengths come into account:
the first one $\delta_0^{af}$ is the domain wall width in the substrate near the
interface, and the second one $\delta_f$ is the domain wall width in the 
ferromagnetic layer.
Since $\delta_f \gg a$, one can neglect the domain wall widening in the ferromagnet.

 Let us consider the order parameter behavior in the antiferromagnet and
the situation at the film-surface interface. Let the atomic step on the interface coincide
with the y-axis of Cartesian coordinate system. For $x \ll -\delta_f$ the conditions
$\theta_{af} = \theta_f = 0$
 are fulfilled, and for $x \gg \delta_f$ the conditions $\theta_{af} =0,\, \theta_f = \pi$ are
met. It follows from the symmetry of the problem that 
$\theta_{af} =0,\, \theta_f = \pi/2$ for $x=0$.
The width of the region at the film-surface interface where the value of $\theta_f - \theta_{af}$
 differs from its optimum ($0$ for $x<0$ and $\pi$ for $x>0$) equals $\delta_0^{af}$. In the
region of $|x|\lesssim \delta_f$ and $|z|\lesssim \delta_f$ there arise
distortions of the order parameter in the substrate (Fig.~\ref{Fig3}).

Analogously to the previous consideration one can estimate the energy $w_1$
inside the film by Eq.~(\ref{w1}) and inside the substrate as
\begin{equation}
 w_1^{af} \approx \frac{\pi}{2}\frac{J_{af} S_{af}^2}{b} \ln \frac{\delta_f}{\delta_0^{af}},
\label{eqn13}
\end{equation}
 The energy $w_2$ is given by Eq.~(\ref{w2}) with $\delta_0^{af}$ instead of $\delta_f$.
After minimization one has
\begin{eqnarray}
\delta_0^{af} &\approx& \frac{1+\alpha_{af}}{\alpha_{af}},\label{eqn12a}\\
 \delta_f &\approx& \gamma a,
\label{eqn12}
\end{eqnarray}
and $w \approx w_1^{af}$, so
the main contribution being caused by the order parameter distortions in the substrate.

 The foregoing estimations relate to the case of well away distances between the
steps. If the distance between the steps decreases and becomes less than the critical
value $R_c = \delta_f$, the domain walls begin to overlap and the film switches to a single
domain state. 
The transition from a polydomain state to a single-domain state is continuous and, in
the strict sense, is not a phase transition.

 If $\gamma \ll 1$, $\alpha_f a \gg 1$ and $\delta_0^f \ll R \ll \delta_f$, 
a static spin vortex arises in the
film near the substrate enclosing the area $z \lesssim R$.

 But if $\gamma \gg 1$, $\alpha_{af} a\gamma \gg 1$, and the value of $R$ is in the range
$\delta_0^{af} \ll R \ll \delta_f$, a similar spin vortex arises 
near the interface in the substrate. Near the interface
spin orientation in the vortex corresponds to the minimum of the
interface energy $w_2$. Far from the interface spin orientation is homogeneous.
The results of modeling are introduced in Fig.~\ref{Fig4}.

For all others cases the distortions of order parameters
are small in both the film and the substrate
and their values can be considered as constants.

 As it was already mentioned, the steps divide the whole interface into the areas
of two types. If the mean magnetization vector makes an angle $\psi$ with the
antiferromagnetic order parameter in the substrate volume, then the value of $\theta_f$
(or $\theta_{af}$) changes from zero to $\psi$ in the vortex occupying the first type area, whereas in
the vortices occupying the second type areas the value of $\theta_f (\theta_{af})$ changes from $\psi$
to $\pi$.

 By an analogy to the ``magnetic proximity'' model \cite{ref7} the energy of the system
can be written down as
\begin{equation}
 W = C_1 \psi^2 + C_2 (\pi - \psi)^2,
\label{eqn14}
\end{equation}
where the ratio of phenomenological constants $C_1$ and $C_2$ is proportional to the ratio
of the areas of both types ($\sigma_1$ and $\sigma_2$ correspondingly). 
If the distributions of these areas in sizes 
for each type are equal, one has
\begin{equation}
 C_i \approx \frac{J_{f(af)} S_{f(af)}^2 \sigma_i}{R b}.
\label{eqn15}
\end{equation}
In the case of $\sigma_1=\sigma_2$,
 without regard for the energy of anisotropy induced by the steps, in both the
vortex phase and the region of small distortions the equilibrium magnetization of the
film must be perpendicular to the antiferromagnetic order parameter. The phase
diagram of the two-layer system is exhibited in Fig.~\ref{Fig5}.

 This specific pattern correlates well with the results of paper,\cite{ref10} where the
``thickness{}--{}vicinal angle $\beta$'' phase diagram for iron film on Cr (001) surface was
investigated. For $\beta$ angles close to zero a polydomain phase was observed for the film
thickness $a<a_c=3.5 nm$. For the film of the critical thickness $a_c$ the characteristic
distance between the edges of random steps fits the value $\gamma a_c$. In thicker films a
single-domain phase was observed, with magnetization perpendicular to the step
edges. The theory proposed above predicts the antiferromagnetism vector to be
parallel to the step edges, its experimental verification being of immediate interest.

 If $\beta \neq 0$, regularly distributed parallel steps are added to random atomic steps.
When the concentration of regular steps becomes dominating $\beta \geq 1^\circ$, the critical
thickness $a_c$ falls down. It follows from our theory that $a_c \propto R \propto t g^{-1}\beta\propto\beta^{-1}$.

 At high values of $\beta$ the orientational phase transition to the phase with
magnetization parallel to the steps takes place.\cite{ref10} This transition is caused by the
anisotropy induced by the steps through relativistic effects, for example, dipole
interaction.\cite{ref13}
\section{Multilayer ``ferromagnet - antiferromagnet'' structure}
Let us consider for simplicity a three-layer system consisting of two
ferromagnetic layers and an antiferromagnetic interlayer. In view of a number of
various parameters, we restrict ourselves to the case of $\gamma \gg 1$ and the layers of equal
thicknesses. Such a system can occur in three different states.
\subsection*{$A$-phase}
 At high values of $R$ all layers are divided into domains with parallel and
antiparallel orientation of magnetizations in ferromagnetic layers. The domain walls
penetrate into all three layers, their coordinates in the layer plane coincide with the
atomic steps edges on each of two interfaces. The magnetizations of ferromagnetic
layers rotate to different sides in the domain wall The antiferromagnetic order
parameter rotates together with the magnetization of that ferromagnetic layer whose
interface is free of a step in the given place.

 The structure and energy of the domain wall depend on the parameter $\alpha_{af}a$.
When $\alpha_{af}a \ll 1$, one can neglect the dependence $\theta_{f(af)}(z)$, i.e. the domain wall
widening (Fig.~\ref{Fig6}a).

The consideration analogous to the case of the film gives the expression (\ref{eqn7})
for the domain wall width in the ferromagnet and 
$\delta_{af}\approx\sqrt{a/\alpha_{af}} = \delta_f/\sqrt{\gamma}\ll\delta_f$
for antiferromagnet one. The main contribution to the domain wall
energy $w_1$ comes from the ferromagnetic layers and $w$ value is given by Eq.~(\ref{eqn8}).

But if $\alpha_{af}a \gg 1$, then, as in the case of a two-layer system,
two length scales come into existence: $\delta_0^{af}=(1+\alpha_{af})/\alpha_{af}$ and $\delta_f'$ which differs
from that given by Eq.~(\ref{eqn12}) because the distortions of the antiferromagnetic order
parameter are now limited by the layer thickness.
The distribution of the order parameters in this wall is shown in Fig.~\ref{Fig6}b.

The domain walls in ferromagnetic layers have the width much higher than $a$,
thus it may be considered as a constant across the layer. In antiferromagnetic interlayer
one can see two regions (Fig.~\ref{Fig6}c). In the first one $|x|\lesssim a$ the
situation is analogous to the two-layer system: $|\nabla \theta| \propto \rho^{-1}$.
The contribution $w_1^{af}$ from the region is 
\begin{equation}
 w_1^{(1)} \approx \frac{\pi}2 \frac{J_{af} S_{af}^2}{b} \ln\frac{a}{\delta_0^{af}}.
\end{equation}
But in the second region $a\ll|x|\ll\delta_f'$ the lines of constant $\theta$ values
are almost parallel to the interfaces. In this region $|\nabla \theta| \propto a^{-1}$.
The contribution of the region to $w_1^{af}$ is
\begin{equation}
 w_1^{(2)} \approx \frac{J_{af} S_{af}^2 \delta_f'}{b a}.
\end{equation}
Taking into account the contribution to $w_1$ from ferromagnetic layers (Eq.~(\ref{w1}))
and the interface energy $w_2$ one comes to the result given by Eq.~(\ref{eqn12a})
for $\delta_0^{af}$,
\begin{equation}
 \delta_f' \approx a \sqrt{\gamma},
\label{eqn16}
\end{equation}
and
\begin{equation}
 w \approx \frac{J_{af} S_{af}^2}{b} \left(\sqrt{\gamma}+\ln\frac{a}{\delta_0^{af}} \right).
\label{eqn17}
\end{equation}
\subsection*{$B$-phase}
 Since the magnetic rigidity of ferromagnetic layers is higher than that of
antiferromagnetic $(\gamma\gg1)$, the transition to a state with practically uniform
ferromagnetic layers, because of the domain walls overlapping, takes place with the
distance $R$ decreasing and reaching the value $R_c=\delta_f(\delta_f')$. An additional energy is
associated with either distortions in antiferromagnetic interlayer or peculiarities of the
structure at the interfaces. Near the Neel temperature of an interlayer $T_N$ ($T_N$ is less
than the Curie temperature of ferromagnet) one has $\gamma \propto T_N/(T_N-T)$, so the
transition $A\to B$ may be performed by heating the system from an initial temperature
$T_0<T_N$. The ``magnetic proximity'' model proposed by Slonczewski \cite{ref7} is
appropriate just in that $R$ range where the $B$-phase exists.

 In the range of distances $\max{\left(a,\delta_0^{af}\right)}\ll R\ll R_c$, the energy dependence
on the angle $\psi$ between the magnetizations of ferromagnetic layers is described by Eq.~(\ref{eqn14}).
Really, for $\alpha_{af} a \gg 1$, in the regions of the first type 
(where parallel orientation of ferromagnetic layers magnetizations is 
energetically favorable) the antiferromagnetic order parameter changes linearly 
with $z$ from one interface to another by the $\psi$ value ($|\nabla \theta| = \psi/a$).
In the regions of the second type 
(where antiparallel orientation of ferromagnetic layers magnetizations is 
energetically favorable) it changes by $\pi-\psi$, $|\nabla \theta| = (\pi-\psi)/a$.
The frustration at interfaces is absent. The boundary energy between regions of the
first and the second types is of no significance in the case $R \gg a$.

In the case $\alpha_{af} a \ll 1$ the antiferromagnetic order parameter
is almost constant, but one has frustrations at the interfaces. The jump of
$\theta$ value equals $\psi/2$ at each interface in the regions of the first type
and $(\pi-\psi)/2$ in the regions of the second type.
The excess energy $w_2$ per two-dimensional unit cell in the regions of the first 
and the second type equals:
\begin{eqnarray}
\tilde w &=& \frac{4 J_{f,af} S_f  S_{af}}{b^2} \sin^2\frac{\psi}{4}, \label{wpsi} \\
\tilde w &=& \frac{4 J_{f,af} S_f  S_{af}}{b^2} \sin^2\frac{\pi-\psi}{4}. \label{wpipsi}
\end{eqnarray}
Eqs.~(\ref{wpsi}-\ref{wpipsi}) are not identical to Eq.~(\ref{eqn14}),
the phenomenological constants  $C_1,C_2$ can be estimated by comparing energy
difference of collinear and $90^o$ orientations in the model proposed here
and in the magnetic--proximity model \cite{ref6}:
\begin{equation}
C_{1,2} = \left\{\matrix{
 \frac{J_{af} S_{af}^2}{2a}\cdot\frac{\sigma_{1,2}}{b^2},\quad \alpha_{af}a \gg 1,\hfill \cr
 \frac{4(\sqrt2-1)}{\pi^2}\; J_{f,af} S_f S_{af} \;\frac{\sigma_{1,2}}{b^2},\quad \alpha_{af}a \ll 1,\hfill
}\right.
\label{eqn18}
\end{equation}
where $\sigma_{1,2}$ are the areas of the first and second types on the layer surface.

 For $\sigma_1=\sigma_2$ the energy minimum is attained at $\psi=\frac{\pi}{2}$, i.e. the
magnetizations of ferromagnetic layers are mutually perpendicular in the absence of an
external magnetic field.

 Ferromagnetic layers are, to be sure, not ideally homogeneous. By analogy to
Ref. \cite{ref14} the change of the quantities $\theta_i,\psi$ across the length of $R$ can be shown to be
of the order of $\left[R/\delta_f(\delta_f')\right]^2\ll 1$.

 Since the radius of the vicinity influencing the value of $\theta_i$ at a given point
equals $\delta_f(\delta_f')$, the deviation of $\theta_i$ and $\psi$ from their mean values is defined by the
fluctuation of the quantity that characterizes the relation between the square areas of
the regions of two types. If the regions of the first type dominate in given vicinity, one
has $\psi<\pi/2$, but if those of the second type, one has $\psi>\pi/2$.

 The characteristic area bounded on by the steps on the layer surface is of the
order of $R^2$, so the number $N$ of the regions in the area $\left[\delta_f(\delta_f')\right]^2$ is of the order of
$\left[\delta_f(\delta_f')/R\right]^2$. When it is assumed that the regions of both types are randomly
distributed the predominance of the regions of one-type measures
$\sqrt{N} \approx \delta_f(\delta_f')/R$. Therefore the characteristic fluctuation of the quantities $\theta_i$ and
$\psi$ is of the order of $R/\delta_f(\delta_f')\ll 1$.

 In the $A$-phase the energy of the system does not depend on the direction of the
order parameter rotation in the domain wall. A different picture is observed in the
$B$-phase: when the walls overlap one another, the degeneration in the rotation direction is
eliminated and a number of metastable states arise. They differ from each other by the
direction and the value of the angle of rotation of antiferromagnetic order parameter in
individual areas bounded by the atomic steps.

 And what is going on with further $R$ decreasing? If $\alpha_{af}a \ll 1$ and $R$ lies in the
range $a\ll R\ll \delta_{af}$, the system arrives at the region of small distortions where the
order parameters are practically uniform, the magnetizations of ferromagnetic layers
being mutually perpendicular and the constants $C_{1,2}$ lowering by a factor of
$\left(R/\delta_{af}\right)^2$ comparing to their estimations (\ref{eqn18}).

\subsection*{$C$-phase}
 Now let us consider the range of small $R$, $R\ll a$. For such a case all
distortions are localized near the interfaces, the interaction between ferromagnetic
layers becomes weak and the main part is played by the energy of interaction between
the neighboring layers demonstrated above with an example of a two-layer system. As
the result, at $\sigma_1=\sigma_2$ the antiferromagnetic order parameter is oriented
perpendicularly to the magnetizations of ferromagnetic layers which thus appear to be
collinear (parallel or antiparallel). Such a state is termed here as $C$-phase.

 For $\alpha_{af}a \gg 1$ and the distance $\delta_0^{af}\ll R\ll a$, the static vortices are
formed in an antiferromagnetic interlayer near the interfaces. For lower values of $R$ the
system changes over to the region of small distortions.

 If $\alpha_{af}a \ll 1$, the transition from the $B$-phase to $C$-phase takes place already
in the region of small distortions. Along with the $B$-phase, the $C$-phase is characterized
by a presence of a wealth of metastable states. The numerical calculations has shown
the $B\to C$ phase transitions to be of the first order. Both phases coexist in the entire
region of $R$ values and their specific energies become equal at a certain value
$R^* \sim a$ (Fig.~\ref{Fig7}). The phase diagram of the three-layer system is exposed in Fig.~\ref{Fig8}. 
In \cite{ref15a} the distribution of spins in a two-layer ferromagnet --- layered 
antiferromagnet system is modeled numerically on the bases of Ising model. However, 
the Ising model corresponds to very strong anisotropy of easy-axis type and is invalid 
for the description of multilayer structures of the Fe/Cr type, for which the anisotropy 
energy in the plane of the layers is much lower than the exchange interaction energy. 
The domain walls formed within the framework of the Ising model have atomic widths, 
so that the unique properties of domain walls in the multilayers ferromagnet --- 
layered antiferromagnet where not observed in the cited paper.
The orientation of spins in the three-layer system has been calculated in Ref. \cite{ref15} within
the framework of a discrete model for the case $R \sim a$, the magnetizations of
ferromagnetic layers being assumed to be mutually orthogonal. The behavior of the
system in the whole range of $R$-values has not been analyzed.

 The value of $R^*$ is independent of temperature, therefore the $B\to C$ phase
transition is not observable under changes of temperature of the system. The transition
from the state with strong biquadratic coupling to the low-temperature state with a
weak coupling between the layers \cite{ref16} is unrelated to the transition considered above.
The transition considered in Ref. \cite{ref16} is caused by the fact that as approaching $T_N$, the
interaction of ferromagnetic layers via the antiferromagnetic order parameter decreases
and becomes equal to the interaction via spin polarization induced in the
antiferromagnet (the RKKY interaction).\cite{ref6} For higher temperature the latter
dominates.
\section{Conclusions}
Let us state the main results and conclusions of the work exposed above.
\begin{enumerate}
\item
 It is proposed a simple model allowing one to find the distribution of spins in
``ferromagnet{}--{}antiferromagnet'' frustrated layered structures.
\item
 The ``thickness{}--{}roughness'' phase diagrams for a thin ferromagnetic film on
antiferromagnetic substrate and for a ``ferromagnet{}--{}antiferromagnet{}--{}ferromagnet''
three-layer system are obtained.
\item
 The transition from a polydomain state to a single-domain state is continuous and, in
the strict sense, is not a phase transition.
\item
 The transition from the phase with mutually perpendicular orientation of
magnetizations of ferromagnetic layers ($B$-phase) to the phase with collinear
magnetizations of the layers ($C$-phase) constitutes a true first-order phase transition.
\item
 The ``magnetic proximity'' phenomenological model proposed by Slonczewski is
shown to be adequate for the $B$-phase only. The parameters of the model are
calculated for the entire range of its applicability.
\end{enumerate}
\acknowledgments
This work is partly supported by
Russian Foundation for Basic Research,
grant 00-02-17162.

 \begin{figure}
 \caption{Frustrations in the system ``ferromagnet{}--{}layered antiferromagnet'' caused by the
presence of atomic steps on the interfaces.}
 \label{Fig1}
 \end{figure}

 \begin{figure}
 \caption{Dependence of the domain wall width on the distance to the interface for $\alpha_{af} a \gg 1$
 ($\alpha_{af} = 1$ and $a =32$).}
 \label{Fig2}
 \end{figure}

 \begin{figure}
 \caption{Domain wall in the two-layer system for $\gamma \gg 1$.
Levels of order parameters $\theta_i$ in radians are shown by various hatching (see insert).
The distribution was found for $\gamma=8$, $\alpha=1$, $a=8$.
The value $z=0$ corresponds to the interface between the film and the substrate.
Step is situated at the point $x=0$.
}
 \label{Fig3}
 \end{figure}


 \begin{figure}
 \caption{Static spin vortices in the film-substrate system for $\gamma \gg 1$.
Levels of order parameters $\theta_i$ in radians are shown by various hatching (see insert).
The distribution was found for $\gamma=8$, $\alpha=0.01$, $a=8$.
The value $z=0$ corresponds to the interface between the film and the substrate.
Steps are situated at the points $x=\pm 10$.
}
 \label{Fig4}
 \end{figure}

 \begin{figure}
 \caption{Phase diagram of the two-layer system. 
 For demonstration the lines corresponding to equations 
$R=\delta_f$ and $R=\delta^{af}_0$ are drawn.}
 \label{Fig5}
 \end{figure}

 \begin{figure}
 \caption{Domain wall in the three-layer system: 
$\alpha_{af} a \ll 1$ (a); $\alpha_{af} a \gg 1$ (b).
Levels of order parameters $\theta_i$ in radians are shown by various hatching (see insert).
The values $z=0$ and $z=16$ correspond to the interlayer interfaces.
Step is situated at the point $x=0$, $z=0$.
The distributions were found for $\gamma=10$, $\alpha=0.01$, $a=16$ (a) 
and $\gamma=10$, $\alpha=1$, $a=16$ (b). 
In the Fig. (c) is represented the central part of the distribution shown in the Fig. (b).
}
 \label{Fig6}
 \end{figure}

 \begin{figure}
 \caption{Dependence of $B$- and $C$-phases specific energy on the distance $R$ between the steps
 ($a=64$, $\alpha_{f}=8$, $\alpha_{af}=1$).}
 \label{Fig7}
 \end{figure}

 \begin{figure}
 \caption{Phase diagram of the three-layer system. 
 For demonstration the lines corresponding to equations 
$R=\delta_f$, $R=a$, and $R=\delta^{af}_0$ are drawn.}
 \label{Fig8}
 \end{figure}
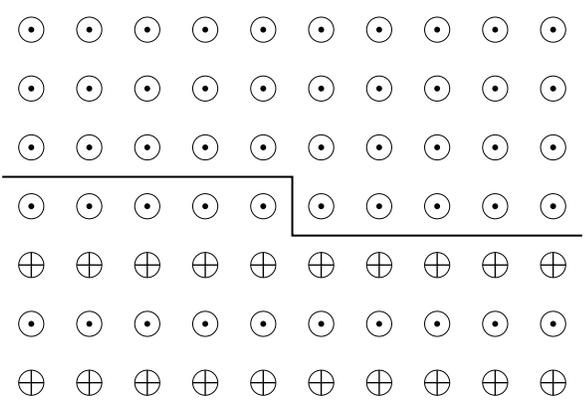
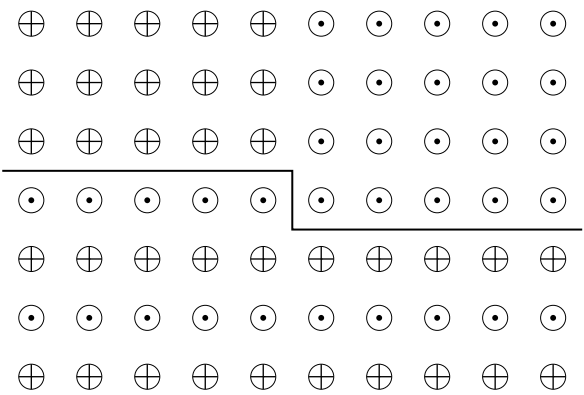
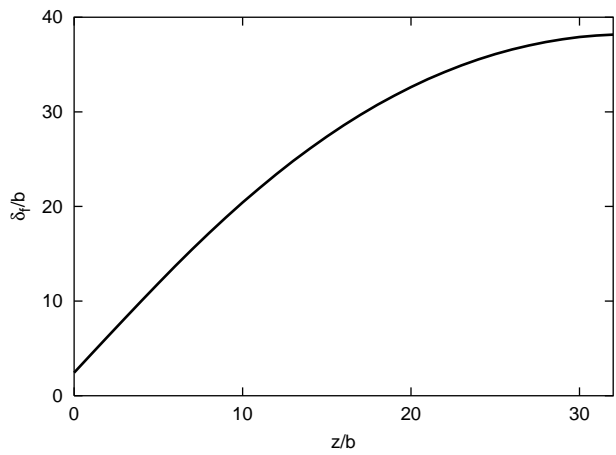
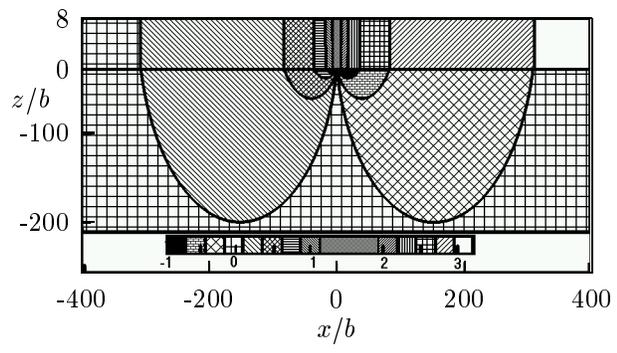
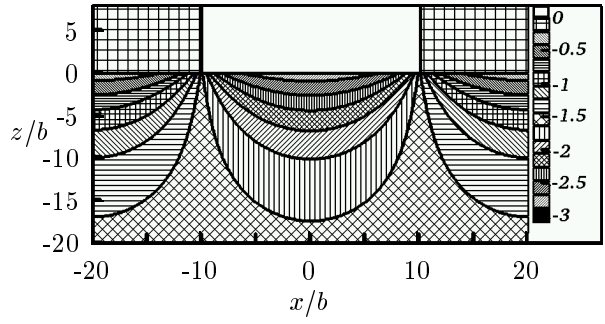
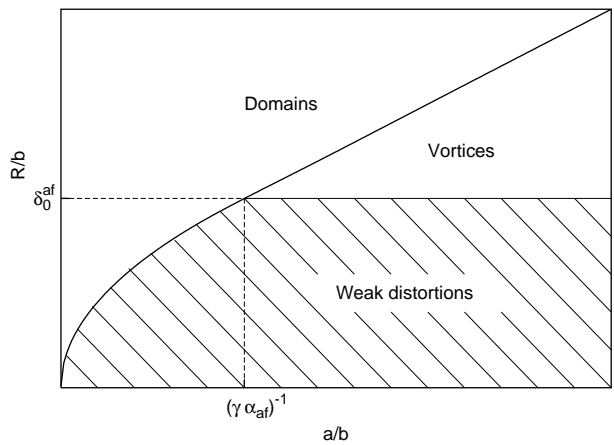
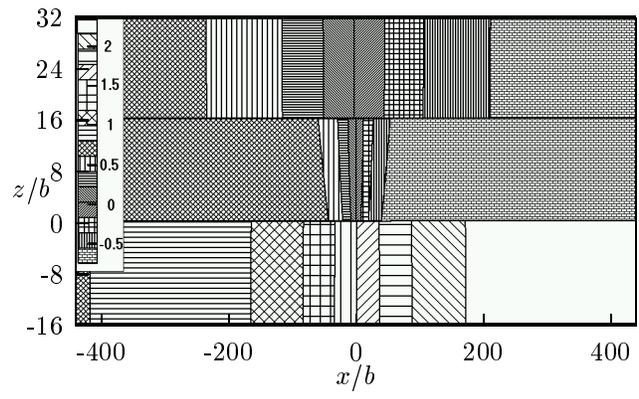
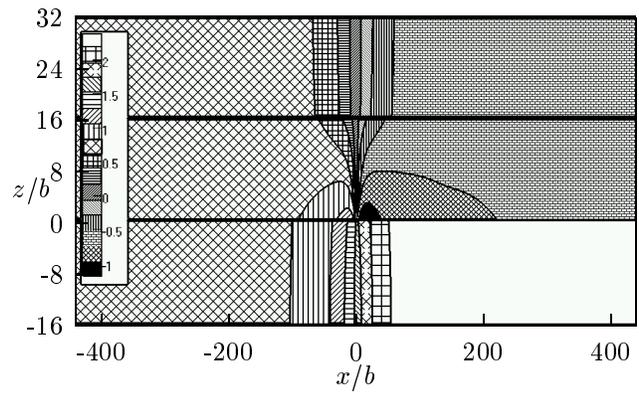
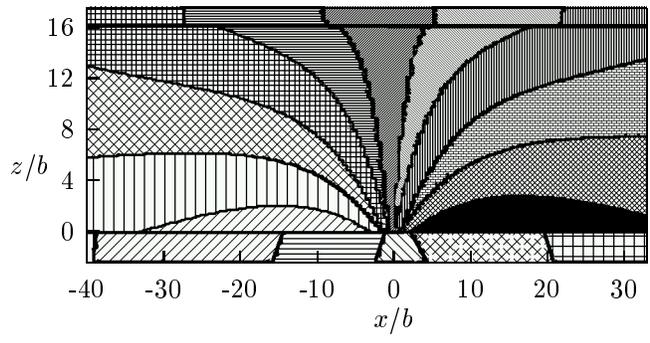
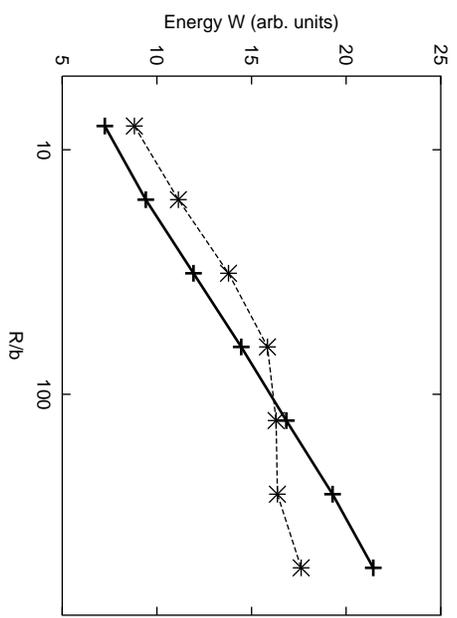
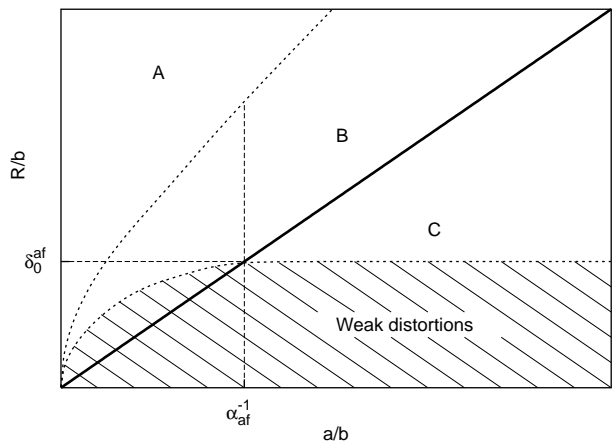
\end{document}